\documentclass[aps,amssymb,prl,twocolumn]{revtex4}
\usepackage{graphicx}
\usepackage{psfrag}

\def\d {{\rm d}}

\def\bn{\begin{itemize}}
\def\en{\end{itemize}}
\def\nm{{\rm nm}}
\def\mm{{\rm mm}}

\def\s{{\rm s}}

\begin{document}
\bibliographystyle{prsty}

\title{Dynamical noise and avalanches in quasi-static plastic flow of amorphous solids}
\author{Ana\"el Lema\^{\i}tre$^{(1)}$}
\author{Christiane Caroli$^{(2)}$}
\affiliation{$^{(1)}$ Institut Navier-- LMSGC, 2 all\'ee K\'epler,77420 Champs-sur-Marne, France}
\affiliation{$^{(2)}$ INSP, Universit\'e Pierre et Marie Curie-Paris 6, Universit\'e Denis Diderot-Paris 7, 
CNRS, UMR 7588 Campus Boucicaut, 140 rue de Lourmel, 75015 Paris, France}

\date{\today}

\begin{abstract}
We build a mean-field model of plasticity of amorphous solids, based on the dynamics of an ensemble shear transformation
zones, interacting via intrinsic dynamical noise generated by the zone flips themselves.
We compare the quasi-static, steady-state properties for two types of noise spectrum:
(G) Gaussian; (E) broad distribution derived from quadrupolar elastic interactions.
We find that the plastic flow proceeds via avalanches whose scaling properties with system size
are highly sensitive to noise tails.
Comparison with available data suggests that non-affine strain fields might
be of paramount importance in the small systems accessible to molecular simulations.
\end{abstract}

\maketitle

Decades of efforts to explain plasticity of amorphous solids, and more generally
of jammed disordered systems (foams, granular media, colloidal glasses\dots),
have converged towards general agreement on the nature of the elementary dissipative events
in these highly multistable systems.
They involve sudden rearrangements of small clusters
comprising a few basic structural units (molecules, bubbles, grains,\dots).
Each 'flip' can be viewed as the shear-induced transformation of an ellipsoidal 
Eshelby inclusion and generates a long-range elastic displacement field with 
quadrupolar symmetry.
In spite of this progress, relating the macroscopic stress response to the 
statistics of elementary events remains a challenging question.

Recently, two phenomenologies of plasticity of jammed systems
(SGR~\cite{Sollich1998} and STZ~\cite{FalkLanger1998}) have been proposed.
For zero-temperature systems, they derive constitutive equations 
based on a few common assumptions for the shear-induced dynamics
of an ensemble of {\it non-interacting\/} shear 'zones' (or 'traps').
Namely, the flippable zones are advected by the external drive at shear rate $\dot\epsilon$
toward their instability threshold; meanwhile, their elastic energy fluctuates
under the effect of a mechanical noise, assumed to be described by a fixed effective temperature.
Flips can thus be triggered before zones reach the instability threshold, with 
{\it strain-rate independent\/}, Arrhenius-like jump probabilities.
In this sense noise, in these models, can be viewed as {\it extrinsic\/}.

Given that noise has a crucial bearing on the form of the resulting constitutive laws,
it would be desirable, however, to relate phenomenological models to microscopic arguments.
Following Argon {\it et al\/}~\cite{ArgonBulatovMottSuter1995}, consider a zone far from its threshold.
As time elapses, its strain is advected by the external drive.
In steady state, flips occurring at random sites 
elsewhere in the material emit--at an average rate  $\sim\dot\epsilon$--elastic 
signals which propagate at sound speed and upon arrival, shift the local strain.
These fluctuating shifts, originating from spatially random sources, 
constitute the {\it dynamical\/}, {\it intrinsic\/}, mechanical noise.
In view of the long range of elastic couplings, one can expect it to give rise,
as in comparable systems with pinning and long range interactions~\cite{KuntzSethna2000},
to avalanches, i.e. 'instantaneous' series of events triggered by a single initial flip.
The existence of correlated flips may have a bearing on the macroscopic plastic properties, 
including the emergence of shear localization.

In this letter, we analyze a mean-field model for the quasi-static dynamics of
a random 2D ensemble of $N$ zones, driven by shear and interacting only via intrinsic,
dynamical noise.
We first clarify the limits of the quasi-static regime in which flips 
can be assumed instantaneous and avalanches are well-separated.
We then construct two versions of the model, differing in their assumptions 
about the noise spectrum: 
model G with Gaussian noise and model E more realistically based upon the quadrupolar 
elastic field~\cite{PicardAjdariLequeuxBocquet2004}.
In both cases, we find, in agreement with previous numerical simulations
of molecular glasses~\cite{MaloneyLemaitre2004,MaloneyLemaitre2006,DTU2006,TanguyLeonforteBarrat2006},
that flips cluster into avalanches of average size $\langle n\rangle$
increasing as a power $\alpha$ of system size $N$.
As expected, the average stress drop decreases with $N$ ensuring proper convergence
towards a smooth stress curve.
The two noise spectra, however, lead to drastic differences in several important features,
including: the average shear stress, the distribution of avalanche sizes 
and its scaling properties, and the scaling exponent $\alpha$.
For model G, simple arguments based upon an approximate master equation 
provide a reasonable prediction for $\alpha$. 
A similar argument fails for model E, probably due to the presence of tails larger that Gaussian,
even though all the moments of the elastic noise are finite.
Confrontation with molecular simulation results points towards the need for further developments,
including in particular non-affine strain field effects.

We consider an ensemble of $N$ identical zones of size $a$, 
randomly distributed with a fixed density $a^2/d^2$, in a 2D elastic medium of lateral size $L$. 
The elastic state of each zone is characterized by an internal strain $\epsilon_i$,
which measures the departure from its zero stress state.
The $\epsilon_i$'s lie below a common stability threshold $\epsilon_c$.
The system is driven by external shear at rate $\dot\epsilon$, which advects all $\epsilon_i$.
When a zone reaches $\epsilon_c$, it disappears and another one is created at an uncorrelated position,
with zero initial stress (hence zero local strain).
The duration of a flip, controlled by acoustic radiative damping, 
is $t_0\sim a/c_s$ with $c_s$ a sound speed.
A flip at site $\vec r_i$ emits an acoustic signal which propagates throughout, 
and modifies the local strains of all other zones $j$, which adjust over a time $\sim t_0$, 
to this space-dependent shift $\delta\epsilon(\vec r_{ij})$. 
Following Picard {\it et al\/}~\cite{PicardAjdariLequeuxBocquet2004}, this elastic field has quadrupolar symmetry, 
hence zero average: it constitutes a noise acting on the $\epsilon_i$.

In steady state, the average flip rate in the whole system is 
$R_{\rm flip}=\delta t^{-1}_{\rm flip}=N\,\dot\epsilon/\epsilon_c$. 
In the absence of avalanches, the noise correlation time is $t_0$, 
and the quasi-static regime corresponds to 
$\delta t_{\rm flip}\gg t_0$ i.e. $\dot\epsilon\ll \dot\epsilon_{\rm flip}=\frac{\epsilon_c}{N\,t_0}$. 
For a glass-like system, with $\epsilon_c\sim 1\%$, $a\sim1\nm$, zone density $a^2/d^2\sim 10^{-1}$, lateral size 
$L\sim 1\mm$, this yields the loose criterion $\dot\epsilon\ll1\s^{-1}$. 
Now, an elastic noise signal may drive some $\epsilon_j$'s beyond $\epsilon_c$, hence trigger secondary flips,
thus initiating an avalanche, whose duration $t_a$ is set by sound propagation.
For a conservative estimate we take the average distance between successive flips to be $L$.
This leads to $t_a=\langle n\rangle\,L/c_s$ with $\langle n\rangle$, the average avalanche size.
The average avalanche frequency is thus 
$R_{\rm a}=\delta t^{-1}_{\rm a}=\frac{N\,\dot\epsilon}{\langle n\rangle\,\epsilon_c}$, and the quasi-static condition becomes~\footnote{Clearly, a statistical description of avalanches also demands that $\langle n\rangle \ll N$. We will
see that this condition is fulfilled in our models.}:
$\dot\epsilon\ll\frac{\epsilon_c\,c_s}{L\,N}=\frac{a}{L}\,\dot\epsilon_{\rm flip}$.
Avalanches drastically reduce the quasi-static range. 
A second important point, usually ignored, is the rapid shrinking of this range with system size.

\paragraph{Models:}
In the quasi-static regime the dynamics reduces to 
steady drift of all $\epsilon_i$'s at fixed  $\dot\epsilon$,
interspersed with instantaneous avalanches. 
We can eliminate time and define our models following a two-step algorithm:
(i) We start from a configuration where all $\epsilon_i<\epsilon_c$.
We identify $\epsilon_{m}={\rm Max}\,({\epsilon_i})$ 
and shift the macroscopic strain, hence all $\epsilon_i$'s, 
by $\Delta\epsilon_{\rm drift}=\epsilon_c-\epsilon_{m}$.
(ii) 
Zone $m$ flips, i.e. is removed while another one is introduced at zero strain.
The resulting elastic signal is modelled by random shifts $\delta\epsilon_i$
of all other $\epsilon_i$'s.
Here, we make the mean-field assumption that the $\delta\epsilon_i$'s are independent 
random variables, which amounts to ignoring the correlations due to the spatial zone arrangement.
The elastic nature of dynamical noise then enters via the spectrum 
of $\delta\epsilon_i$'s (see below).
This first flip yields a new configuration $\epsilon_i^{(1)}$.
If $\epsilon_m^{(1)} < \epsilon_c$ we are back to step (i).
Otherwise, an avalanche starts: we count the number $z=q_1$ of zones that flip at this stage 
and treat their noise signals successively.
If the first signal triggers $q_2$ new flips, $z$ is updated as $z\to z+q_2-1$, etc.
The avalanche stops when $z=0$; its size $n=1+\sum q_i$.
We implement this algorithm with two specifications of dynamical noise. 

Model E is based upon the results of Picard {\it et al\/}~\cite{PicardAjdariLequeuxBocquet2004}:
the elastic strain at position $(r,\theta)$ from a flip is:
$\frac{2}{\pi}\,\frac{a^2\,\Delta\epsilon_0}{r^2}\,\cos(4\theta)$, 
where $\theta$ is measured from a principal axis of the macroscopic strain,
and the elementary strain $\Delta\epsilon_0/\epsilon_c={\pi}\,{L^2}/{2N\,a^2}$.
Each instance  of the noise $\delta\epsilon_i$ is obtained by uniform Monte-Carlo sampling 
of $(r,\theta)\in]d,L[\times]-\pi,\pi[$, where $d$
is the average distance between nearest zones and $L^2=N\,d^2$.
Note the size dependence of this elastic noise, of variance
$\Delta_N={2\,A_0\epsilon_c^2}/{N}$, where, in our circular geometry, the factor $A_0=1/\pi^2$.
One easily checks that the noise distribution  $\rho_{\rm E}(\delta\epsilon)$
exhibits a $1/\delta\epsilon^2$ behavior up to the cut-off 
$2a^2\Delta\epsilon_0/\pi\,d^2$ beyond which it vanishes, so that all its moments are finite.

Model G assumes that noise is simply Gaussian with the same variance $\Delta_N$.
It will provide a test of the influence of tail shapes in $\rho$.

\begin{figure}[b]
\psfrag{nav}{{\Large $\log\langle n\rangle$}}
\psfrag{sys_size}{{\Large $\log N$}}
\includegraphics[width=0.45\textwidth,clip]{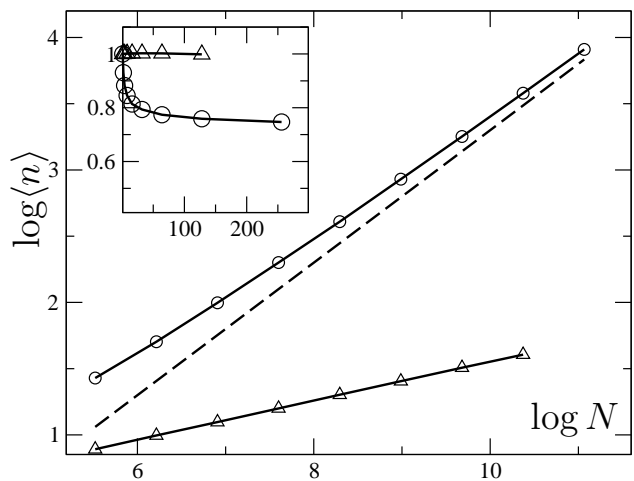}
\caption{\label{fig:1}
Average avalanche size $\langle n\rangle$ vs $N$ for models G (triangles) and E (circles).
The dashed line has slope $1/2$.
Insert: $\langle n\rangle N^{-\alpha}/\langle n\rangle(N_0) N_0^{-\alpha}$ vs $N/N_0$ with $N_0=250$. 
}
\end{figure}

\paragraph{Numerical results:}
In order to study the statistical properties of such models in steady state, 
we eliminate the initial transient ($\epsilon<2\,\epsilon_c$) and perform ergodic 
averaging over long time intervals involving $\gtrsim10^5$ avalanches.
We plot on Figure~\ref{fig:1} the variations of the average avalanche size $\langle n\rangle$ with 
system size $N$. The data for model E fit closely the power law 
$\langle n\rangle_E\sim N^{\alpha_E}$, with $\alpha_E=0.147$, over the whole range (two decades).
The data for model G are consistent with an asymptotic (see insert)
power law behavior $\langle n\rangle_G\sim N^{\alpha_G}$, with the trivial exponent $\alpha_G=1/2$.

\begin{figure}
\psfrag{epsilon}{{\Large $\epsilon/\epsilon_c$}}
\psfrag{figa}{{(a)}}
\psfrag{figb}{{(b)}}
\includegraphics[width=0.45\textwidth,clip]{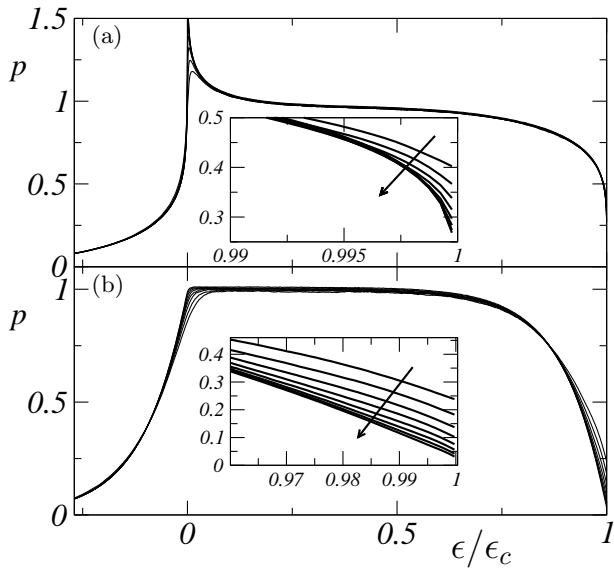}
\caption{\label{fig:2} Steady state distribution $p$ of zone strains vs $\epsilon/\epsilon_c$
for:
(a) model E; (b) model G. Inserts: blow-ups of the near-threshold region. 
The arrows indicate increasing values of $N=250\times2^q$ ($q=0,\ldots,7$).
}
\end{figure}

We also compute the ergodic average of the macroscopic stress $\sigma$. 
Since the macroscopic
strain equals the average zone strain, $\sigma=2\,\mu\,\langle\epsilon_i\rangle$ 
(with $\mu$ the shear modulus).
We find for $\beta=\sigma/(2\,\mu\,\epsilon_c)$ the values $\beta_G=0.40$ and $\beta_E=0.42$.
The 5\% difference, though small, is numerically significant.
We also check, in agreement with the above power law scalings, that stress fluctuations 
decrease rapidly with system size, the ranges of the noisy stress curves becoming
completely separated for $N>32000$.
Different $\beta$ values imply differences in the stationary distribution $p$ of $\epsilon_i$.
$p(\epsilon)$ is represented on Figure~\ref{fig:2} for both models and various values of $N$.
For each model, $p$ converges rapidly almost everywhere towards a limit curve,
which explains that the $\beta$ values vary by less than $0.4\%$ (model G)
and $0.02\%$ (model E) when $N$ increases from $1000$ to $32000$.
While $p_G$ and $p_E$ are similar in most of the $\epsilon$ range, they present
significant differences in two regions, $\epsilon\sim0$ and $\epsilon\sim\epsilon_c$.
The peak in $p_E$ results from refeeding zones at $\epsilon=0$ after flips.
The larger an avalanche, the more the corresponding peak is broadened by ulterior
flips within the avalanche itself.
We therefore attribute the washing out of the peak for model G to the fact that it exhibits
much larger avalanches.
More significant for our purpose is the detailed behavior of $p$ near threshold,
which reflects avalanche statistics.
In particular, $p_c=p(\epsilon_c)$ is directly related to $\langle n\rangle$.
Indeed, the average flip and avalanche rates verify $R_{\rm flip}=R_{\rm a}\,\langle n\rangle$.
On the other hand, the {\it advected\/} zone flux in steady state reads 
$f_{\rm adv.}=\dot\epsilon\,p_c\,N$.
It comprises all zones initiating avalanches
(while the ``descendants'' of each of these mother zones 
feed the noise contribution to the total flux $R_{\rm flip}$).
Hence $f_{\rm adv.}=R_{\rm a}$, from which $p_c=(\langle n\rangle\,\epsilon_c)^{-1}$.
Since the data indicate that $\langle n\rangle$ diverges for $N\to\infty$,
they also indicate that the absorbing boundary condition $p_c=0$ should
hold asymptotically here.
This relation provides a consistency test of our calculations.
We determine $p_c$ with the help of a second order polynomial 
extrapolation near $\epsilon_c$, with sampling intervals 
$\Delta\epsilon/\epsilon_c=1.25\times10^{-6}$.
We find that the relation $p_c\langle n\rangle\,\epsilon_c=1$ holds within 1\% for model G,
and 3\% for model E.

\paragraph{Discussion}
In order to try and obtain analytical estimates for the above avalanche size scalings,
we describe the evolution of $p(\epsilon)$ by the approximate master equation:
\begin{equation}
\frac{\partial p}{\partial t}=-\dot\gamma\,\frac{\partial p}{\partial \epsilon}
+\int_{-\infty}^{\epsilon_c}\,\d\epsilon'\,p(\epsilon')\,w(\epsilon-\epsilon') - \Gamma\,p(\epsilon)
+\frac{f}{N}\,\delta(\epsilon)
\label{eqn:master}
\end{equation}
where $w$ is the single flip transition probability: 
$w(\delta\epsilon)=N\,\dot\epsilon\,\rho(\delta\epsilon)/\epsilon_c$
with $\rho$ the above-defined noise distribution.
Here $\Gamma=\int_{-\infty}^\infty\d\delta\epsilon\,w(\delta\epsilon)$.
The delta term, proportional to the normalized zone flux $f/N=\dot\epsilon/\epsilon_c$, 
accounts for post-flip reinjection and ensures the conservation of zone number.
In this approximation, advection operates between all single flips, 
which amounts to neglecting intra-avalanche time correlations.
This leads to an average advective $\epsilon$-shift during an avalanche 
$\Delta\epsilon_{\rm adv.}\sim\dot\epsilon/R_{\rm a}\sim\epsilon_c\,\langle n\rangle/N$,
to be compared with the average diffusive broadening 
$\Delta\epsilon_{\rm diff.}\sim\sqrt{\langle n\rangle\,\Delta_N}$ with 
$\Delta_N\sim\dot\epsilon^2/N$ the noise variance. 
Hence, $\Delta\epsilon_{\rm adv.}/\Delta\epsilon_{\rm diff.}\sim\sqrt{\langle n\rangle/N}
\sim N^{-(1-\alpha)/2}$ which suggests that our approximation should improve in the large $N$ limit.
Integration of equation~(\ref{eqn:master}) in steady state yields:
$$
\frac{f}{N}=\frac{\dot\epsilon}{\epsilon_c}=\dot\epsilon\,p_c+\int_{-\infty}^{\epsilon_c}\d\epsilon\,
p(\epsilon)\,\int_{\epsilon_c}^\infty\d\epsilon'\, w(\epsilon-\epsilon')
$$
Since $w$ is peaked around zero, we expand $p(\epsilon)$ close to $\epsilon_c$ to first order:
$p(\epsilon)\sim p_c+(\epsilon-\epsilon_c)\,p_c'$. Using $p_c=(\langle n\rangle\,\epsilon_c)^{-1}$,
we obtain for the average avalanche size:
\begin{equation}
\langle n\rangle = \left[1+\frac{N\,\langle \delta\epsilon\rangle_+}{\epsilon_c}\right]
\left[1+\frac{N\,p_c'\,\langle \delta\epsilon^2\rangle_+}{2}\right]^{-1}
\label{eqn:scaling}
\end{equation}
where the (semi)-moments
$\langle \delta\epsilon^r\rangle_+=\int_0^\infty\d(\delta\epsilon)\,w(\delta\epsilon)
\,\delta\epsilon^r$.
For both model, $\langle \delta\epsilon^2\rangle_+=A_0\,\epsilon_c^2/N$, while
$\langle\delta\epsilon\rangle_{+}^{(G)}=\epsilon_c\,\sqrt{\frac{A_0}{\pi\,N}}$ and
$\langle\delta\epsilon\rangle_{+}^{(E)}=\epsilon_c\,\sqrt{\frac{8\epsilon_c}{\pi}\,\frac{\log N}{N}}$. 
If $p_c'$ converges towards a finite value $p_c'^{(\infty)}$, equation~(\ref{eqn:scaling})
predicts that, for large systems,\\
- for model G:~~~~ $\langle n\rangle\sim N^{1/2}$\\
- for model E:~~~~ $\langle n\rangle\sim \log N$\\
While this prediction accounts satisfactorily for the numerical results for model G, 
we have checked (see also Fig.~\ref{fig:1}) that the log scaling is ruled out by our data.
The reason for this failure is illustrated by the insert of Fig.~\ref{fig:2}-(a).
For model E, we find that, for increasing $N$, $p$ becomes increasingly steep in the near
vicinity of $\epsilon_c$. In the $N$-range investigated, we see a marked, non-saturating,
increase of $|p_c'|$, suggesting a possible divergence, higher derivatives increasing even faster.
This highly singular behavior clearly invalidates the above truncated expansion of $p$ for model E.
Conversely, for model G, we find numerically that $p'_c\,\epsilon_c^2\,A_0$ 
does exhibits convergence towards $\simeq -1.3$.

\begin{figure}
\psfrag{pdf}{{\Large $\varpi(n)$}}
\includegraphics[width=0.5\textwidth,clip]{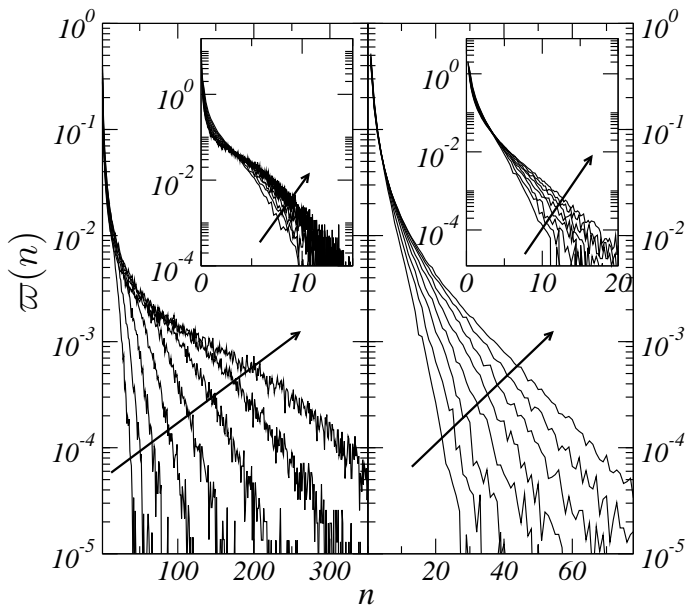}
\caption{\label{fig:3} 
Distribution $\varpi(n)$ of avalanche sizes, for model G (left) and E(right), 
and for all system sizes. 
Inserts: $\langle n\rangle\,\varpi(n)$ vs $n/\langle n\rangle$.
The arrows indicate the direction of increasing sizes.
}
\end{figure}

This discussion points towards our central result:
the slowly space decrease typical of elastic interactions crucially affects avalanche behavior.
This holds not only for their average size, but, as well for their size distribution $\varpi(n)$.
As appears on Fig.~\ref{fig:3}, for both models, $\varpi(n)$ 
decreases exponentially for $n\gg\langle n\rangle$. However, a fundamental difference
appears when attempting to rescale 
$\varpi(n)$ as $\varpi(n) = (1/\langle n\rangle)\,f(n/\langle n \rangle)$ 
(inserts of Fig.~\ref{fig:3}).
For model G, the quality of the collapse, good for $n\gtrsim\langle n\rangle$ at all $N$, 
improves with system size in the small $n/\langle n\rangle$ range.
On the contrary no such scaling holds for model E.

These results on dynamical noise effects, which remain preliminary,
indicate two routes for further investigations.

Our mean-field approximation wipes out from the start the directional effects arising 
from the quadrupolar structure of the elementary events, which are responsible for the 
preferential avalanche orientations observed in the LJ glass simulations 
by Maloney {\it et al\/}~\cite{MaloneyLemaitre2004,MaloneyLemaitre2006} 
and Tanguy {\it et al\/}~\cite{TanguyLeonforteBarrat2006}.
In order to evaluate the robustness of the mean-field
scalings and also to start addressing the issue of localization,
full simulations of model E will be necessary.

The second series of questions arises when confronting our results with 
those of~\cite{MaloneyLemaitre2004,MaloneyLemaitre2006}.
Maloney {\it et al\/} find in the quasi-static regime that
$\langle N\rangle$ scales approximately as $\sqrt{N}$. Moreover their avalanche distribution
obeys quite well the above-mentioned scaling. Analogous results emerge from the 3D simulation
of an amorphous metallic alloy by Bailey {\it et al\/}~\cite{DTU2006}.
This seems to plead in favor of Gaussian noise.
This a priori surprising conclusion raises a further question.
Recently, Leonforte {\it et al\/}~\cite{LeonforteBoissiereTanguyWittmerBarrat2005}
have shown that the elastic response of amorphous solids
self-averages into the continuum elastic response only beyond a length scale $\xi$
of order $\sim20$ molecular diameters $a_0$.
For $r<\xi$, the elastic response is dominated by non-affine effects.
So, for the existing molecular simulations focusing on avalanche dynamics,
where $L$ is limited to $\lesssim 50\,a_0$ units, 
noise tails are very likely to be controlled by elastic non-affinity.
Incorporating such effects into models of intrinsic noise thus appears
as an important, though intricate, issue.


\end{document}